\documentclass[12pt]{article}
\usepackage{graphicx}
\usepackage{pstricks,pst-node}
\usepackage{pst-node,pst-text,pst-3d} % Used by prosper etc.
\usepackage{verbatim}
\usepackage{tabularx}
\usepackage{amssymb,epsfig}
\input epsf.tex 
\newcommand{\beq}{\begin{equation}}
\newcommand{\eeq}{\end{equation}}

\newcommand{\beqn}{\begin{eqnarray}}
\newcommand{\eeqn}{\end{eqnarray}}

\font\af=msbm12

\def\C{{\af C}}

\def\be{\begin{eqnarray}}    
\def\ee{\end{eqnarray}}
\def\Dsl{\,\raise.05ex\hbox{/}\mkern-9.5mu D}
\def\mbox#1#2{\vcenter{\hrule \hbox{\vrule height#2in 
\kern#1in \vrule} \hrule}} 
\def\boxeqn#1{\vcenter{\vbox{  \hrule height2pt \hbox{\vrule
width 2pt \kern3pt\vbox{\kern3pt
\hbox{${\displaystyle #1}$}\kern3pt}\kern3pt\vrule width 2pt}\hrule height2pt}}}
\def\tr{{\rm \, tr\, }}
\def\back{{{\raise.4em\hbox{$\, _\backslash\,$}}}}

 2

\let\spec=    b
\def\frac#1#2{{#1\over #2}}

\def\big R{{\hbox{{\bigfield R}}}}
\def\bbig R{{\hbox{{\bbigfield R}}}}

\font\af=msbm10
\font\afs=msbm8
\font\afm=msbm12

\def\C{\hbox{\afm C}}
\def\Cc{\hbox{\afs C}}
\def\N{\hbox{\af N}}

\def\C{\hbox{\af C}}

\def\N{\hbox{\af N}}

\def\I{\hbox{\af I}}

\def\I{\hbox{\afm I}}

\mathchardef\imath="717B
\def\inbar{\,\vrule height1.5ex width.4pt depth0pt}
\def\IB{\relax{\rm I\kern-.18em B}}
\def\IC{\relax\hbox{$\inbar\kern-.3em{\rm C}$}}
\def\ID{\relax{\rm I\kern-.18em D}}
\def\IE{\relax{\rm I\kern-.18em E}}
\def\IF{\relax{\rm I\kern-.18em F}}
\def\IG{\relax\hbox{$\inbar\kern-.3em{\rm G}$}}
\def\IH{\relax{\rm I\kern-.18em H}}
\def\II{\relax{\rm I\kern-.18em I}}
\def\IK{\relax{\rm I\kern-.18em K}}
\def\IL{\relax{\rm I\kern-.18em L}}
\def\IM{\relax{\rm I\kern-.18em M}}
\def\IN{\relax{\rm I\kern-.18em N}}
\def\IO{\relax\hbox{$\inbar\kern-.3em{\rm O}$}}
\def\IP{\relax{\rm I\kern-.18em P}}
\def\IQ{\relax\hbox{$\inbar\kern-.3em{\rm Q}$}}
\def\IR{\relax{\rm I\kern-.18em R}}
\font\cmss=cmss10 \font\cmsss=cmss10 at 10truept%!!! should be 7pt
\def\IZ{\relax\ifmmode\mathchoice
{\hbox{\cmss Z\kern-.4em Z}}{\hbox{\cmss Z\kern-.4em Z}}
{\lower.9pt\hbox{\cmsss Z\kern-.36em Z}}
{\lower1.2pt\hbox{\cmsss Z\kern-.36em Z}}\else{\cmss Z\kern-.4em Z}\fi}
\def\IGa{\relax\hbox{${\rm I}\kern-.18em\Gamma$}}
\def\IPi{\relax\hbox{${\rm I}\kern-.18em\Pi$}}
\def\ITh{\relax\hbox{$\inbar\kern-.3em\Theta$}}
\def\IOm{\relax\hbox{$\inbar\kern-3.00pt\Omega$}}

\def\CB{{\cal B}}

\def\CH{{\cal H}}
\def\CL{{\cal L}}

\relax 
\title{Unital Positive  Maps and Quantum States}

\author{
M. Asorey\\{\footnotesize\it Departamento de F\'\i sica Te\'orica. Facultad de Ciencias}.\\
{\footnotesize\it  Universidad de Zaragoza, 50009 Zaragoza. Spain}\\[2ex]
 A.  Kossakowski \\{\footnotesize\it Institute of Physics, Nicolaus Copernicus University}\\
{\footnotesize\it Tor\'un 87-100, Poland}\\[2ex] 
G. Marmo\\{\footnotesize\it  Dipartimento  di Scienze Fisiche}\\
{\footnotesize\it Universit\'a
Federico II di Napoli
and  INFN, Sezione di Napoli}\\
{\footnotesize\it  Complesso Univ. di Monte Sant' Angelo, Via Cintia, 80125 Napoli, Italy}\\[2ex] 
E.C.G. Sudarshan\\{\footnotesize\it Department of Physics. University of Texas at Austin}\\
{\footnotesize\it Austin, Texas 78712-1081} }
\begin{document}
\maketitle

\begin{abstract}

We analyze the structure of the subset of states generated by unital completely 
positive quantum maps, A witness that certifies that a state does 
not belong to the subset generated by a given map is constructed. We analyse
the representations of positive maps and their relation to 
quantum Perron-Frobenius theory.

\end{abstract}


\hspace{12pt}{\it PACS}: {\small 03.65.Bz, 03.67.-a, 03.65.Yz}\\
\vspace{0pc}
\hspace{24pt}{\it Keywords}:\small \  Positive maps,  Perron-Frobenius theory\newline
\maketitle

\section{Introduction}

A  genuine feature of quantum physics is  the existence of relations between dynamical maps and 
quantum states. The analysis of such relations  permits to relate  
entanglement properties of quantum states of  composite systems and  positivity properties
of quantum maps. The interest on positive maps  arises from its broad spectrum  of   mathematical  and physical
applications \cite{Sudarshan61}--\cite{effrostormer}.
Although, from the viewpoint of dynamical maps (quantum channels)
most of the interest has  focused on completely positive maps, because
of their physical interest, positive maps which are not completely positive have  also revealed  
very useful as entanglement witness. Their classification of positive maps which are not
completely positive is a challenging problem which still remains
open.

In general,  positive maps which are not completely positive  might be useful as 
discriminators between quantum states
which belong to subsets included into the images of completely positive maps.
The aim of the present paper
is to further explore the relations between quantum states
and dynamical maps  from this perspective. In particular, we use the Perron-Frobenius theory
to show that all eigenvalues $\lambda$ of unital positive maps remain inside  to the unit disk
$|\lambda|\leq 1$.

In the abstract approach to quantum mechanics a physical system is characterized by 
an unital $\C^\ast$--algebra $A$,  c.f. \cite{emch}. In the algebra $A$ there is a  distinguished 
 cone $A^+$ generated by positive elements of $A$,
 i.e. elements which can be written in  the form $a^\ast a$, $a\in A$.
The unit $e$ is a very special element of $A^+$.
 The set of all states $S(A)=S$ 
is the convex set of normalized ($\omega(e)=1$) linear continuous functionals $\omega:A\to \C$
which are positive,  i.e.  $\omega(a)\geq 0$ for all $a\in A^+$.

A linear map $\varphi:A\to A$ is called self-adjoint if $\varphi(a^\ast)=\varphi(a)^\ast$,  positive if 
$\varphi:A^+\to A^+$, and unital if $\varphi(e)=e$.

Unital positive maps $\varphi$ on $A$ are also called  dynamical maps \cite{Sudarshan61}. Any dynamical map  $\varphi$
 define, by duality, an affine map in the space of states $\varphi^\ast:S\to S$, 
 i.e., $\omega(\varphi(a))=\varphi^\ast(\omega)(a)$, for all $a\in A$ 
and $\omega\in S$.

The interest on positive maps in $\C^\ast$-algebras arises 
 for different reasons \cite{Sudarshan61}--\cite{effrostormer}:

i)  unital positive maps are generalizations of   states, $\ast$--homomorphisms, Jordan homomorphisms, and
conditional expectations,

ii) dual maps corresponds to some physical operations which can be performed on the physical systems (measurement,
time evolution),

iii) there exists a strong relation between the classification of entanglement and positive maps \cite{horodecki3}

Let $M_k(A)$ be the algebra of $k \times k$ matrices whose entries are elements of $A$ and $M^+_k(A)$ be the positive
cone in $M_k(A)$. For $k\in \N$ and $\varphi: A\to A $ we define the maps $\varphi_k:M_k(A)\to M_k(A)$ and 
$\bar{\varphi}_k:M_k(A)\to M_k(A)$ by $\varphi_k([a_{ij}])=[\varphi(a_{ij})]$ and $\bar{\varphi}_k([a_{ij}])=[\varphi(a_{ji})]$.
A map $\varphi$ is said to be $k$-positive ($k$-copositive) if the map $\varphi_k$ ($\bar{\varphi}_k$) is
positive. A map is $\varphi$ is said to be $k$-decomposable if $\varphi_k$ and $\bar{\varphi}_k$  are positive for
matrices $[a_{ij}]\in M_k(A)^+$ with transposes  $[a_{ji}]$ also in  $M_k(A)^+$.

A map $\varphi:A\to A $ is said to be completely positive, completely copositive and decomposable if  it is 
 $k$-positive,  $k$-copositive and $k$-decomposable for all $k\in \N$, respectively.
According to  St\o rmer's  theorem  \cite{stormer82b} any decomposable map $\varphi:A\to \CB(\CH)$ into
the $\C^\ast$-algebra of bounded operators $\CB(\CH)$  on  a complex Hilbert $\CH$, can be decomposed as the sum
$\varphi(a)=\varphi_1(a)+\varphi_2(a)$, of a  completely positive  $\varphi_1$ and a
 completely copositive $\varphi_2$ maps.

A positive map which is not decomposable is called  indecomposable. In the case $A=M_n(\C)$ every
decomposable map $\varphi:M_n(\C)\to M_n(\C)$ can be written in the form
\beq
\varphi(a)=\sum_k a_k^\ast a a_k+\sum_l b_l^\ast a^T b_l,
\eeq
where $a_1,a_2,\cdots; b_1,b_2,\cdots \in M_n(\C)$. It is well known \cite{stormer63,woronovicz76a, gorini76a}
that for two-dimensional systems every positive map $\varphi:M_2(\C)\to M_2(\C)$ is decomposable. 
The first example of an indecomposable map in $M_3(\C)$ was given by 
Choi \cite{choi75b,choi80} for three-dimensional systems (see \cite{terhal}--\cite{ha03} for  generalizations).

The structure of positive maps in $M_d(\C)$ for $d\geq 3$ is not well known, except for  completely positive, completely
copositive  and decomposable maps. It is interesting to note that in the case of 
unital positive projections $\pi: A\to A$, i.e. maps  $\pi: A^+\to A^+$ with
$\pi(e)=e$ and $\pi^2=\pi$, positivitiy properties can be characterized in terms of the properties of the image
$\pi(A)$, i.e. without referring to ancilla systems \cite{choieffros,tomiyama,effrostormer,stormer80}.

Let us assume that $A$ is an unital $\C^\ast$-algebra and  $\pi: A\to A$ is a unital positive projection.
 It is well know \cite{choieffros,tomiyama} that the image $\pi(A)$ is a $\C^\ast$-subalgebra of $A$
under the product
\beq
\pi(a)\circ \pi(b)=\pi(ab)
\eeq
if and only if $\pi$ is completely positive, %and $\pi$ is automatically completely positive if $\pi(A)$ is a $\C^\ast$-subalgebra of A. 
Similar results hold for decomposable unital projections. Unfortunately, the problem of similar
characterization of positive unital maps is still open.

\section{Quantum states and Unital Positive Maps}

In quantum information theory  (see e.g. \cite{albert}) certain classes of quantum states play a relevant 
 role, e.g. separable, PPT, NPT, etc. These states span convex subsets of the set of physical states.

An interesting feature is the existence of relations between some classes of quantum states and
 physical operations which can be performed 
on a physical system. The simplest construction that we shall analyze in this paper can be formulated  in 
terms of unital positive maps.

Let $A$ be a unital $\C^\ast$-algebra, and $S=S(A)$ the set of quantum states,  i.e. continuous, linear, 
positive and normalized functionals on $A$.
Let us assume that $\varphi:A\to A$ is a unital positive map and let $\varphi^\ast S\to S$ be its dual map.
% By duality, $\varphi$ defines an affine  map $\varphi^\ast:S\to S$, given by
% \beq
% \varphi^\ast(\rho)(a)=\rho(\varphi(a))
% \eeq
% for any $\rho\in S$ and $a\in A$.
We  can associate to any  positive unital map  $\varphi$ on $A$  the
 subset of states given by  $\varphi^\ast(S)\subseteq S$.

In terms of $\varphi^\ast(S)$ one can define a new cone $A^+(\varphi)$ in $A$:
\beq
A^+(\varphi)=\biggl\{a=a^\ast\in A; \rho(a)\geq 0\  {\rm for \ all}\  \rho\in\varphi^\ast(S)\biggr\}.
\eeq
It is clear that $A^+\subseteq A^+(\varphi)$. Let us assume that $\varphi^\ast(S)\subset S$,  i.e. 
$A^+\subset A^+ (\varphi)$, and let $\rho\in S$ be an state which does not belong to the subset $\varphi^\ast (S)$,  i.e. 
$\rho\notin \varphi^\ast (S)$. Then,  there exists
$a_\rho\in A^+(\varphi)$ and $a_\rho\notin A^+$ such that
\beq
\rho(a_\rho)<0
\eeq
The element $a_\rho\in A^+(\varphi)$ plays a role of witness that $\rho\notin \varphi^\ast (S)$.

In the case where $\varphi$ is a unital positive projection $\pi$ on $A$,  i.e., $\pi(e)=e, \, \pi^2=\pi$, the structure
 of the cone $A^+(\pi)$ is very simple. One has $A^+(\pi)=A^++({\I}-\pi)(A)=0$ since
$A=\pi(A)+(\I-\pi)(A)$ and $\pi({\I}-\pi)(A)=0$. In this case the image $({\I}-\pi)(A)$
is the set of witnesses for the set $\bar{\pi^\ast(S)}$. A state $\rho\in S$ does not belong to $\pi^\ast(S)$ iff
there exist an element $a\in ({\I}-\pi)A$ such that $\rho(a)\neq 0$ since for any   $\sigma \in \pi^\ast(S)$, $\sigma (a)=0$
for any $a\in ({\I}-\pi)A$.

It is commonly accepted that only completely positive unital maps are related to  physical 
operations which can be performed on a physical system cf. \cite{kraus} (see also counterexamples
in \cite{shaji}). Consequently the  subsets  of states that are   generated by completely positive unital maps are
physically relevant.
It is, thus,  interesting to provide some examples of subsets generated by completely positive unital maps.

Let $\CH$ be a complex finite dimensional  Hilbert space with $\dim \CH=d$, $M=M(\CH)$ the $\C^\ast$-algebra
of linear operators on $\CH$,  and $M^+\equiv M(\CH)^+$ the  cone of positive operators in $M=M(\CH)$.

The set of quantum states  $S=S(M)$ can be identified with the set of all density operators of $\CH$, by the relation $\rho(a)=\tr (\rho a)$.

Let $p=(p_1,\cdots,p_N)$ be a family of projection operators on mutually orthogonal subspaces of $\CH$ such that
$p_1+\cdots+p_N=\I$, and $\omega=(\omega_1,\cdots,\omega_N)$ a family of states 
$\omega_1,\cdots,\omega_N\in S$ such that
\beq
\label{dt}
\tr (\omega_\alpha p_\beta)=\delta_{\alpha \beta};\qquad \alpha, \beta=1,\cdots,N
\eeq
It is easy to verify that the map $\pi(p,\omega):M(\CH)\to M(\CH)$ defined by the relation
\beq
\label{dtt}
\pi ( p,\omega)(a)=\sum_{\alpha=1}^N p_\alpha \tr (\omega_\alpha a)
\eeq
is a completely positive unital projection on $M(\CH)$, and its dual $\pi^\ast(p,\omega):S\to S$
has the form 
\beq
\pi^\ast ( p,\omega)(\rho)=\sum_{\alpha=1}^N \omega_\alpha \tr (p_\alpha \rho).
\eeq
Let us observe that invariant states of $\pi^\ast(p, \omega)$,  i.e. those states  $\rho\in S$
such that $\pi^\ast(p, \omega)(\rho)=\rho$, have the form
\beq
\rho=\sum_{\alpha=1}^N c_\alpha\, \bar{\omega}_\alpha,
\eeq
where $c_1,\cdots,c_N\geq 0$, $ \sum_{\alpha=1}^N c_\alpha=1$.

Choosing $\omega_\alpha=\widetilde{p}_\alpha=p_\alpha/\tr p_\alpha$, the relations 
(5) are automatically satisfied and the corresponding projections (6) will be
denoted by $\pi(p)$,  i.e.,
\beq
\label{cuss}
\pi(p)(a)=\sum_{\alpha=1}^N p_\alpha \tr (\tilde{p}_\alpha a).
\eeq
Moreover, $\pi^\ast(p)=\pi(p)$. The subset of $S$ generated by $\pi(p)$,  i.e. the subset
$\pi^\ast(p)(S)$ is given by
\beq
\pi^\ast(p)(S)=\biggl\{\sum_{\alpha=1}^N p_\alpha \tr (\tilde{p}_\alpha \rho); \rho\in S\biggr\}.
\eeq
Let us now consider the case $\CH=\C^{n}\otimes\C^{n}$, with an orthonormal basis 
$e_1,\cdots,e_n$ in $\C^n$, and  $e_{ij}=e_i(e_j.\cdot)$ in $\C^{n}\otimes\C^{n}$. Define the orthogonal projections
in $\C^{n}\otimes\C^{n}$
\beq
p'_1=\frac1{n}\sum_{ij=1}^n e_{ij}\otimes e_{ij}
\eeq
\beq
p'_0=\I_n\otimes\I_n-p'_1
\eeq
and the corresponding $\pi(p')$, then the subset $\pi^\ast(p')(S)$ is the class of Horodecki isotropic states \cite{horodecki}.
On the other hand the orthogonal projections
\beq
p''_0=\frac1{2} (\I_n\otimes\I_n+F)
\eeq
\beq
p''_1=\frac1{2} (\I_n\otimes\I_n-F)
\eeq
where
\beq
F=\sum_{ij=1}^n e_{ij}\otimes e_{ji}
\eeq
is the flip operator, define another unital projector $\pi(p")$. 
The set $\pi^\ast(p'')(S)$ coincides with the set of Werner states \cite{werner}.

It is easy to check that all classes of multipartite states constructed in \cite{chrus,chrus1} are generated by 
completely positive unital projections $\pi(p)$ with appropriate choice of $\CH$ and projectors $p_1,\cdots, p_N$.
There exists a second class of completely positive unital projections which are  of the form
\beq
\label{cus}
\pi(a)=\sum_{\alpha=1}^N p_\alpha a p_\alpha 
\eeq
and correspond  to a measurement process. These maps are self-dual,  i.e. $\pi^\ast=\pi$, and in the case that
$p_1,\cdots, p_N$ are one-dimensional, then (\ref{cus}) reduces to (\ref{cuss}).

Let us consider the completely positive unital map $\varphi:M(\CH)\to M(\CH)$ such that
\beq
\varphi(a)= \frac1{d-1}a+\biggl(1-\frac1{d-1}\biggr) \I_d \tr (\omega_0 a)
\eeq
where $\omega_0=\I_d/d$ is the maximally mixed state.

The map $\varphi$ is self-dual,  i.e. $\varphi^\ast=\varphi$, and it is easy to verify that
\beq
\varphi^\ast(S)=\biggl\{\rho\in S; \tr \rho^2\leq \frac1{d-1}\biggr\}.
\eeq
On the other hand the cone $M^+(\varphi)$ defined by the set $\varphi^\ast(S)$, i.e.
\beq
M^+(\varphi)=\biggl\{a=a^\ast \in M; \tr (a \rho)\geq 0\ {\rm for\ all}\ \rho \in \varphi^\ast (S)  \biggr\}.
\eeq
has the following structure
\beq
M^+(\varphi)=\biggl\{a=a^\ast \in M; \tr a \geq 0,\ \tr a^2\leq (\tr a)^2  \biggr\} \supset M^+.
\eeq
since $\varphi^\ast (S)$ is a ball.

In the case $\CH=\C^d$ the set $\varphi^\ast(S)$ has been considered in Ref. \cite{kossakowski}
while in the case $\CH=\C^{d_1}\otimes\cdots\otimes\C^{d_k}$ with $d=d_1,\cdots,d_k$,
 $\varphi^\ast(S)$ is the largest separable ball around the maximally mixed state \cite{gurvits}.

\section{Representations of linear maps}

Let $\CL(M_d,M_d)$ be the vector space of linear maps $\varphi:M_d\to M_d$, and 
$M_d\otimes M_d$  the vector space of linear maps $\widehat{\varphi}:\C^d\otimes\C^d\to 
\C^d\otimes\C^d$. Given two orthonormal bases  $\{f_\alpha\}$ and $\{g_\alpha\}$ in $M_d$,
the relation
\beq
\label{bone}
\CL(M_d,M_d)\ni \varphi \longrightarrow\widehat{\varphi}=
\sum_{\alpha=1}^{d^2}\,  f_\alpha\otimes \varphi(g_\alpha)
\eeq
defines
an  isomorphism between $\CL(M_d,M_d)$ and $M_d\otimes M_d$, c.f. \cite{arrighi}--\cite{kimura}.
The isomorphism is an isometry which maps  the hermitian product
%maps the hermitian product $(\cdot,\cdot)$  is defined by
\beq
(\varphi,\psi)=\sum_{\alpha=1}^{d^2}\, \tr_{\Cc^n}[\varphi(f_\alpha)^\ast \psi(f_\alpha)]
\eeq
of $\CL(M_d,M_d)$ into that
\beq
<\widehat{\varphi},\widehat{\psi}>= \tr_{\Cc^d\otimes\Cc^d}[\widehat{\varphi}^\ast \widehat{\psi}].
\eeq
of $\C^d\otimes\C^d$, i.e.
\beq
(\varphi,\psi)=<\widehat{\varphi},\widehat{\psi}>
\eeq

It is easy to show that the maps
\beq
\label{bb}
\gamma_{\alpha\beta}(a)=f_\alpha \, a g_\beta^\ast,\quad \alpha,\beta= 1,2,\cdots,d^2
\eeq
and
\beq
\epsilon_{\alpha\beta}(a)=f_\alpha \, \tr\ ( g_\beta^\ast\, a),\quad \alpha,\beta= 1,2,\cdots,d^2
\eeq
define two orthonormal bases of $\CL(M_d,M_d)$, i.e.
\beq
(\gamma_{\alpha\beta},\gamma_{\mu\nu})=\delta_{\alpha \mu}\, \delta_{\beta \nu}
\eeq
and
\beq
(\epsilon_{\alpha\beta},\epsilon_{\mu\nu})=\delta_{\alpha \mu}\, \delta_{\beta \nu}.
\eeq
Any  map $\varphi\in \CL(M_d,M_d)$ can be written as
\beq
\varphi(a)=\sum_{\alpha,\beta=1}^{d^2}\,A_{\alpha\beta}\,\gamma_{\alpha\beta}(a)
\eeq
 or 
\beq
\varphi(a)=\sum_{\alpha,\beta=1}^{d^2}\,B_{\alpha\beta}\, \epsilon_{\alpha\beta}(a),
\eeq
where
\beq
\label{bbb}
A_{\alpha\beta}=(\gamma_{\alpha\beta},\varphi), \quad
B_{\alpha\beta}=(\epsilon_{\alpha\beta},\varphi).
\eeq

Using (\ref{bb})-(\ref{bbb}) one finds  the relations 
\beq
\label{aas}
A_{\alpha\beta}=\sum_{\mu,\nu=1}^{d^2}\,\tr\, (f_\alpha^\ast f_\mu g_\beta g_\nu^\ast)\,  B_{\mu\nu}
\eeq
and 
\beq
\label{bes}
B_{\alpha\beta}=\sum_{\mu,\nu=1}^{d^2}\,\tr\, (f_\alpha^\ast f_\mu g_\beta g_\nu^\ast)\, A_{\mu\nu}.
\eeq

A special case is provided by the choice $f_\alpha=g_\alpha=e_{ij}$, where $e_{ij}= e_i(e_j,\cdot)$ are
the projectors defined by an orthonormal basis $e_i,\cdots,e_d$ of $\C^d$. In this case 
the correspondence (\ref{bone})  is 
known as Jamiolkowski isomorphism \cite{jamiolkowski} and the corresponding matrices
$A_{ij,kl}$, $B_{ij,kl}$ were introduced by Sudarshan \cite{Sudarshan61}, while the relations
(\ref{aas})(\ref{bes}) have the form
\beq
A_{ij,kl}=B_{ik,jl}.
\eeq

On the other hand choosing $f_\alpha=g_\alpha$ yields the following representations for $\varphi$,
\beq
\label{aaes}
\varphi(a)=\sum_{\alpha=1}^{d^2}\,A_{\alpha\beta}\, f_{\alpha}\, a \, f_{\beta}^\ast
\eeq
 or 
\beq
\label{bees}
\varphi(a)=\sum_{\alpha=1}^{d^2}\,B_{\alpha\beta}\, f_{\alpha}\, \tr\,  ( f_{\beta}^\ast a). \,
\eeq

The advantage of using the representation (\ref{aaes}) is due to the fact  
that  $A_{\alpha\beta}=\bar{A}_{\beta\alpha}$ for   any selfadjoint 
map $\varphi$ , i.e. a map which satisfies that $\varphi(a^\ast)=\varphi(a)^\ast$, and
$ [ A_{\alpha\beta}] $ is a semi-definite positive matrix if
$\varphi$ is a completely positive map \cite{kraus}.

On the other hand, the composition of maps in the representation (\ref{bees}) corresponds to the
product of $B$ matrices, while the completely positivity condition takes the form
\beq
\sum_{\alpha,\beta=1}^{d^2}\,B_{\alpha\beta}\, f_{\alpha}^T \otimes  f_{\beta}^\ast\geq 0.
\eeq 
Let us now consider the eigenvalue problem for the map $\varphi:M_d\to M_d$, i.e.
\beq
\label{eigen}
\varphi(a)=\lambda\, a.
\eeq
In the basis $\{f_\alpha;\alpha=1,\cdots d^2\}$, the matrix $a$ can be written as
  $$a=\sum_{\alpha=1}^{d^2}a_\alpha f_\alpha$$
 and using (\ref{bees}) one gets that
(\ref{eigen}) is equivalent to
\beq
\label{eigenn}
\sum_{\beta=1}^{d^2} B_{\alpha\beta} a_\beta =\lambda a_\alpha.
\eeq
For a positive map $\varphi$, the eigenvalue problem (\ref{eigen}) (\ref{eigenn}) is related to the quantum
version of Frobenius theory \cite{frobenius}\cite{evans}, and one has the following result.

{\underline {\bf Theorem}}

If $\varphi:M_d\longrightarrow M_d$ is positive all its eigenvalues satisfy the inequality
\beq
|\lambda|\leq \parallel \varphi(\I_d)\parallel_\infty,
\eeq
where $\parallel \cdot\parallel_\infty$ is the operator norm in $M_d$.

{\underline {\it Proof}}

Since $\varphi$ is positive it follows from the theorem by Russo and Dye \cite{Russo} (see also \cite{paulsen}) that 
$\parallel\varphi\parallel_\infty=\parallel\varphi(\I_d)\parallel_\infty$ and, then
\beq
|\lambda|\parallel a\parallel_{_\infty}=\parallel \varphi(a)\parallel_{_\infty} \leq \parallel a\parallel_{_\infty} \parallel\varphi\parallel_{_\infty}=\parallel a\parallel_{_\infty}\parallel\varphi(\I_d)\parallel_{_\infty},
\eeq
which proves the theorem.

{\underline {\it Corollary}}\\ 
If $\varphi$ is positive and unital, i.e. $\varphi(\I_d)=\I_d$,  all eigenvalues of $\varphi$
lie in the unit circle.

Let $\varphi$ be a completely positive unital map on $M_d$. Then, there exist two  bi-orthonormal bases $\{f_\alpha\}$
$\{g_\alpha\}$ in $M_d$ such that
\beq
\varphi(a)=\sum_{\alpha=1}^{d^2}\, \lambda_\alpha\, f_{\alpha}\, \tr\,  ( g_{\alpha} a),
\eeq
where
\beq
\tr (f_\alpha g_\beta)=\delta_{\alpha\beta}
\eeq
\beq
|\lambda_\alpha|\leq 1,\qquad \alpha,\beta =1,\cdots, d^2
\eeq
and
\beq
\sum_{\alpha=1}^{d^2}\, \lambda_\alpha\, f^T_{\alpha}\, \otimes  g_{\alpha}\geq 0,
\eeq
such that
\beq
\varphi(f_\alpha)= \lambda_\alpha\, f_{\alpha}
\eeq
and
\beq
\varphi^\ast(g_\alpha)= \lambda_\alpha\, g_{\alpha}
\eeq
$\varphi^\ast$ being the dual map.

Let us assume that $\varphi^\ast(\omega)=\omega$, i.e. $\omega$ is an invariant state. Then $\varphi$
has the following representation
\beq
\varphi(a)=\sum_{\alpha=1}^{d^2}\, \lambda_\alpha\, f_{\alpha}\, \tr\,  ( g_{\alpha} a),
\eeq
in terms of the bi-orthonormal bases $\{f_\alpha;\alpha=1,\cdots, d^2\}$ and $\{g_\alpha;\alpha=1,\cdots, d^2\}$ of $M_d$
given by
\beq
  g_{d^2}=\omega,\quad f_{d^2}=\I_d 
\eeq
and
\beq
g_\alpha=h_\alpha,   f_{\alpha}=h^\ast_\alpha-\I_d \tr(\omega h^\ast_\alpha)\qquad  \alpha=1,\cdots, d^2-1,
\eeq
where $\{h_\alpha\in M_d; \alpha=1,\cdots, d^2-1\}$ is a family of orthonormal traceless matrices, i.e.
\beq
 \tr h_\alpha=0;\, \tr\, h_\alpha h^\ast_\beta=\delta_{\alpha\beta}.
\eeq
% with  $$, and  
%$$
 %$sl(d, \C)$.

To illustrate the theorem let us consider the following example of  unital map $\varphi: M_d\to M_d$ given by
\beq
\varphi(a)=\sum_{i,j=1\atop i\neq j}^{d-1}\, \alpha_{j-i}\, e^\ast_{ij} a e_{ij} +\alpha_0 \sum_{i, j=1}^{d-1}\, \beta_{ij}\, e^\ast_{ii} a e_{jj} 
\eeq
where  the indices $j-i$ are understood  mod $d$, i.e. if $j<i,   \alpha_{j-i}:=\alpha_{d-j+i}$, and the coefficients $\alpha_i$
satisfy the  constraints
\beq
%\nolabel
\alpha_0,\alpha_1,\cdots, \alpha_{d-1}\geq0, %\qquad \alpha_i = \alpha_{d-i}
\eeq
and
\beq
%\nolabel
\sum_{ j=0}^{d-1}\, \alpha_{j}=1.
\eeq
The matrix $\beta=[\beta_{ij}]\geq 0$ is positive with $\beta_{ii}=1$ for all  $i =0,\cdots,,d-1$, which implies that
all non-diagonal entries  verify the inequality $|\beta_{ij}|\leq 1$.

It is easy to show that $\varphi$ is a completely positive bi-stochastic map and  satisfies the following
relations
\beq
\varphi(e_{ij})=\alpha_0 \beta_{ij} e_{ij} \qquad i\neq j
\eeq
\beq
\varphi(u_m)=\rho_m u_m,\qquad m=0,\cdots,d-1,
\eeq
 where
\beq
u_m=\frac1{\sqrt{d}}\sum_{ j=0}^{d-1}\, \lambda^{jm}e_{jj},
\eeq
\beq
\tr(u_m u^\ast_n)=\delta_{mn}
\eeq
\beq
\lambda={\rm exp }\left(\frac{2\pi}{d}i\right)
\eeq
and
\beq
\rho_m=\sum_{ j=0}^{d-1}\, \alpha_{j} \lambda^{-jm}.
\eeq
Indeed, the eigenvalues of $\varphi$: $\rho_0,\cdots,\rho_{d-1}$ and $\alpha_0\beta_{ij}, i\neq j=0,\cdots,d-1$
are contained in  the unit disk. The map $\varphi$ can also be represented in the form
\beq
\varphi(a)=\sum_{ j=0}^{d-1}\, \rho_{j} u_j\, \tr (u_j^\ast a)+\alpha_0 \sum_{i,j=0\atop i\neq j}^{d-1}\, \beta_{ij}\, e_{ij}\, \tr (e^\ast_{ij}a)   
\label{fin}
\eeq
in terms of the orthonormal basis 
$\{f_{mn}, m,n=0 \cdots d-1; f_{mm}=u_m, f_{mn}=e_{mn}  (m\neq n)\}$ 
of 
$M_d$,
which  provides the spectral decomposition of the map $\varphi$.

\section*{Acknowledgements}
The work of M.A. and G. M. has been partially supported by a cooperation grant INFN-CICYT.
 M.A. has also been partially supported by the Spanish MCyT grant FPA2006-02315 and DGIID-DGA 
(grant 2007-E24/2). A. K. has been supported by MNiSW grant No. NN202300433.

%\section*{References}

\bigskip

\end{document}